\documentclass[11pt,letterpaper]{article}
\usepackage[letterpaper,bindingoffset=0.2in,%
            left=1in,right=1in,top=1in,bottom=1in,%
            footskip=.25in]{geometry}
            
\usepackage{amsmath}
\usepackage{amssymb}

\usepackage[hidelinks]{hyperref}

\usepackage{lineno}

\usepackage{color}

\usepackage{microtype}
\DisableLigatures[f]{encoding = *, family = * }

\usepackage{setspace}

\usepackage{graphicx}

\usepackage{lineno}
\usepackage[font=small,format=plain,labelfont=bf,up,textfont=up]{caption}

\usepackage[numbers]{natbib}

\usepackage{abstract}

\begin{document}


\begin{center}
{\Large
\textbf{\textcolor{black}{
Multiple stable states in microbial communities \\explained by the stable marriage problem
}}
}
\\
\vskip 10pt 
\normalsize
Akshit Goyal$^{1,\dagger}$, Veronika Dubinkina$^{2, \dagger}$, Sergei Maslov$^{2,\ast}$
\\
\vskip 10pt 
$^1$ \emph{Simons Centre for the Study of Living Machines, National Centre for Biological Sciences, \\ Tata Institute of Fundamental Research, Bengaluru 560 065, India.}
\\
\vskip 5pt
$^2$ \emph{Department of Bioengineering and Carl R. Woese Institute for Genomic Biology,\\
University of Illinois at Urbana-Champaign, Urbana, IL 61801, USA.}
\\
\vskip 10pt
\small
$\dagger$ These authors contributed equally to this work.
\\
$\ast$ Correspondence: \href{mailto:maslov@illinois.edu}{\texttt{maslov@illinois.edu}}
\end{center}

\section*{Abstract}
\noindent
Experimental studies of microbial communities routinely reveal that they have \textcolor{black}{ 
multiple} stable states. While each of these states is generally resilient, \textcolor{black}{certain perturbations such as antibiotics, probiotics and diet shifts, result in transitions to other states.
Can we reliably both predict such stable states as well as direct and control transitions between them?} Here we present a new conceptual model --- inspired by the stable marriage problem \textcolor{black}{in game theory and economics --- in which microbial communities naturally exhibit multiple stable states,
each state with a different species' abundance profile.}
Our model's core ingredient is that microbes utilize nutrients one at a time while competing with each other.
Using only two ranked tables, \textcolor{black}{one with} microbes' nutrient preferences and \textcolor{black}{one with their} competitive abilities, we can determine all \textcolor{black}{possible stable states as well as predict inter-state transitions, triggered by the removal or addition of a specific nutrient or microbe.
Further,} using an example of 7 \emph{Bacteroides} species common to the human gut utilizing 9 polysaccharides, we predict that mutual complementarity in nutrient preferences enables these species to coexist at high abundances. 

\section*{Introduction}
\noindent
One of the major goals of microbiome research is to achieve a mechanistic understanding of the structure, function, and dynamics of microbial communities \cite{Konopka2009,Konopka2015}. The recent rapid proliferation of metagenomics and other -omics data has promoted correlation-based, large-scale statistical analyses of these ecosystems \cite{Xochitl2015}. One common property revealed by these studies is that communities can often exist in multiple or alternative stable states, distinguished from each other by differences in the abundance profiles of surviving species. Examples of this include the human gut microbiome \cite{Lozupone2012, Faith2013}, bioreactors \cite{Zhou2013}, and soil communities\cite{Fukami2011}. Moreover, external perturbations --- such as the temporary introduction (or removal) of nutrients (or microbes) --- can trigger  transitions between these stable states. This is often the basis for the effect of prebiotics and probiotics on the gut microbiome \cite{Costello2012, David2014} and disturbances in bioreactors or other engineered environments \cite{Briones2003}. However,  our ability to predict stable states as well as direct and control their transitions remains limited. Developing a deeper conceptual understanding of community structure, we believe, is an important step towards such an endeavor.

Ever since pioneering theoretical work by MacArthur and Tilman \cite{MacArthur1961,Tilman1982}, resource competition has been a promising approach to modeling stable states in microbial communities.  Following 
Ref. \cite{MacArthur1961}, contemporary models of microbial communities typically assume that microbes simultaneously co-utilize several substitutable nutrients as sources of carbon and energy \cite{Klitgord2010, Coyte2015, Goldford2017, Advani2017, Tikhonov2017, Wingreen2017}. However, as first described by Monod \cite{Monod1949}, many microbes tend to utilize these nutrients in a specific sequential order. 
When exposed to a mixed medium containing multiple nutrients, microbes begin to grow by first utilizing their most preferred one. Upon the exhaustion of this nutrient, and after a period of stasis known as the lag phase, they undergo a diauxic shift and resume growth using the next available nutrient down in their hierarchy \cite{Monod1949}. 
This continues until all consumable nutrients in the medium that the microbe could grow on are exhausted. 

Recent work by Martens and collaborators \cite{Martens2017, Martens2015, Martens2012} has established that many species in \textit{Bacteroides} (the most prevalent genus in the human gut microbiome \cite{Qin2010, HMP2012}) exhibit this kind of preferential nutrient utilization --- with respect to polysaccharides present in a typical diet \cite{Flint2008}. Interestingly, even species such as \textit{B. ovatus} and \textit{B. thetaiotaomicron} --- which are closely-related evolutionarily --- display rather different  polysaccharide preference hierarchies \cite{Martens2017}. \textcolor{black}{Similar results have also been demonstrated for \emph{Bifidobacterium} species \cite{Riviere2018}.} In addition, many of these 
species are simultaneously present in the gut at high abundances. This is in spite of their similar nutrient utilization capabilities \cite{Martens2015, Raghavan2015} that should have promoted competition and mutual exclusion \cite{Hardin1960}. This apparent `habitat filtering' --- where potential metabolic competitors are frequently detected together at high abundances --- remains a puzzling observation.  

Describing community dynamics where microbes utilize nutrients one at a time can be approached either via mechanistic or conceptual models. To develop mechanistic models however, the main obstacle is that they rely on the knowledge of a large number of quantitative parameters, e.g. growth curves of individual microbes, kinetic rates of adsorption and release of small molecules, etc. The vast majority of these parameters are hard to measure and are currently unknown. This further necessitates the need for conceptual models with a much more coarse-grained description of interactions between microbes and nutrients. In particular, the first question that such models need to deal with concerns `matching': how do complex communities divide resources among their constituent microbes?



\textcolor{black}{In this study we present a new conceptual modeling approach that provides mechanistic insights into several phenomena in microbial communities, specifically: the existence of multiple stable states and inter-state transitions, as well as restructuring and resilience of these states. Our model is inspired by a decades-old economics work: the stable marriage or stable allocation problem, developed by Gale and Shapley in the 1960s \cite{Gale1962} and awarded the Nobel prize in economics in 2012. We also apply this approach to predict patterns in polysaccharide utilization preferences of 7 \emph{Bacteroides} species residing in the human gut. We believe that our model can help bridge the gap between statistical analyses based on metagenomic data and a detailed predictive description of community dynamics.}

\section*{Results}
\subsection*{A model of microbial community dynamics inspired by the stable marriage problem}
{\color{black}The traditional formulation of the Stable Marriage Problem (SMP) is the following: $N$ men and $N$ women have to be matched pairwise in $N$ `marriages'. Every person has associated with them a preference list of all members of the opposite sex, ranked from their most preferred marriage partner (rank $1$) to their least preferred one (rank $N$). A matching is `stable' if it has no `blocking pairs', i.e. it has no man-woman pair (who are not currently married to each other) who would both prefer each other to their current marriage partners. One can show that stability with respect to blocking pairs is sufficient to ensure stability with respect to a coalition of any size \cite{Gusfield1989}.
Gale and Shapley proved \cite{Gale1962} that there is always at least one such stable matching, and introduced a `men-proposing' algorithm to find it. According to this algorithm every men first proposes to his top choice partner. If a woman receives more than one proposal, she temporarily accepts the most suited partner according to her preference list and rejects the others. Men rejected during the first round propose to their second choice and so on. If a woman later on  receives a proposal that is better than her current partner, she accepts it and releases her previous choice. One can prove that the state achieved at the end of this men-proposing procedure is stable \cite{Gusfield1989}. In general there are many different stable states for a given set of preference lists (on average $(N/e)\text{log}N$ for random lists but occasionally exponentially many more). When the set of men and women have unequal sizes, the number of pairs in any matching is given by the size of the smaller set. Furthermore, in all stable states, the partners left without spouses are always the same \cite{Gusfield1989}. Another version of the problem is one with unacceptable partners (partial lists). In this case, one can show that the number of pairs in a stable matching is generally smaller than the number of men and women. As in the previous case, the same set of partners are left without spouses in every stable state \cite{Gusfield1989}.
The stable marriage problem still remains a field of active mathematical research. In particular, some of the recent work addresses various aspects and extensions of the original problem, such as the notion of `universal beauty' and correlations in preference lists \cite{Caldarelli2000}, scaling behaviors \cite{Dzierzawa2000}, partial information \cite{Zhang2001}, three-dimensional preferences and agents \cite{Knuth1997} and versions with ties in preference lists \cite{Miyazaki2015}.}

In our application \textcolor{black}{of this problem to microbial communities,} a set of `marriages' constitutes a one-to-one pairing between microbial species and substitutable nutrients. Consider a set of microbes capable of utilizing the same set of fully substitutable nutrients (e.g. carbon/energy sources).
A more general case when each microbe could utilize only a subset of all available nutrients (incomplete ranked lists) is discussed later on in our study.

\begin{figure}
\centerline{\includegraphics[width=\linewidth]{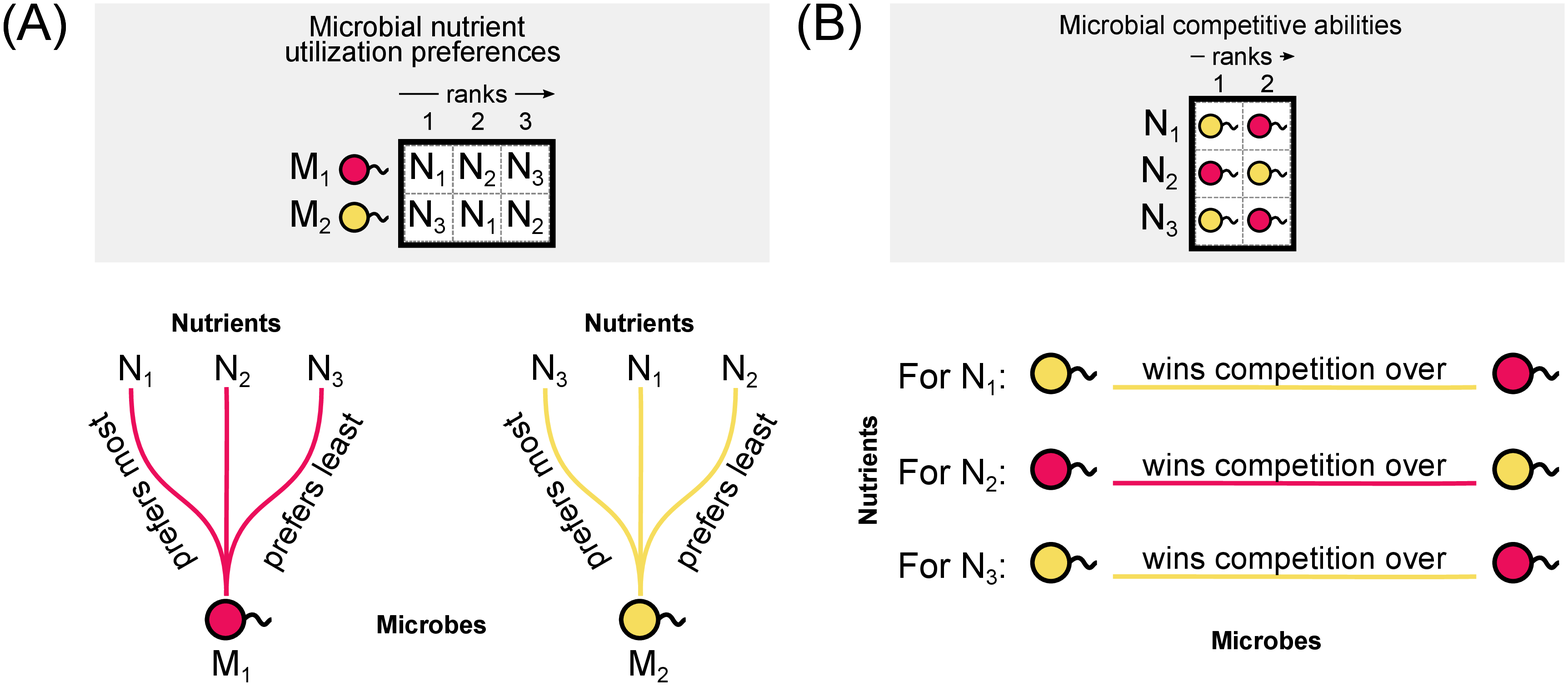}}
\caption{\textbf{Ranked interaction tables encode microbes' nutrient preferences and competitive abilities.}
\\ Two ranked tables with each microbe's preferences towards nutrients (panel A) and their competitive abilities with respect to each particular nutrient (panel B) fully define our 
model. We illustrate them using 2 microbial species, $M_1$ and $M_2$, represented correspondingly by red and yellow circles, and 3 nutrients, $N_1$, $N_2$ and $N_3$. Both species can use all three nutrients. \textbf{(A)} Microbial nutrient preferences: the red species prefers nutrient $N_1$ the most (rank 1 in the table above), $N_2$ next (rank 2), and $N_3$ the least (rank 3), while the yellow species prefers nutrients in the order: $N_3 > N_1 > N_2$. \textbf{(B)} Microbial competitive abilities: the red species (rank 1 
) can displace the yellow species (rank 2 
) in a competition for utilizing the nutrient $N_2$, 
but will be displaced by the yellow species when competing 
for nutrients $N_1$ and $N_3$. 
}
\label{keyfig0}
\end{figure}

%
%
The central assumption in our model is that every microbe consumes these nutrients  
in a diauxic (or more generally polyauxic) fashion\textcolor{black}{, i.e. 
one nutrient after another in a specific sequential order.
This order} 
is encoded in microbe's transcriptional regulatory network combined with diverse post-transcriptional mechanisms of catabolite repression \cite{Gorke2008, Deutscher2008}.
Detailed kinetic modeling of catabolite repression in even one organism (\emph{E. coli}) is 
rather complicated and involves up to 63 state variables connected up to 473 kinetic parameters, 
\emph{most} of which are not known experimentally \cite{Kremling2015}. 
The advantage of the SMP-based approach is that it depends only on the 
ranked microbial preferences towards nutrients, thus bypassing the need 
for precise measurements of such kinetic parameters. These ranked preferences 
ranging from $1$ (the most preferred nutrient, such as glucose for \emph{E. coli}) to 
$N$ (the least preferred one) are illustrated in figure \ref{keyfig0}(A) 
and may be different even between closely related microbial species \cite{Martens2017}. 

If two or more microbes attempt to simultaneously consume the same nutrient, 
we refer to this event as competition, whose outcome is determined by the 
relative competitive abilities of the respective microbes. In our model, 
the competitive ability of a microbe on a given nutrient is in direct proportion 
to the rate at which it uptakes this nutrient from the medium. Thus the 
microbe with the largest uptake rate would drive that nutrient to the 
lowest extracellular concentration, thereby preventing other microbes 
from growing on it \cite{Tilman1982}. The SMP approach requires only the 
knowledge of a ranked table of microbial competitive abilities ranging 
from $1$ (the most competitive microbe for a particular nutrient) to $M$ 
(the least competitive out of $M$ microbes) (see figure \ref{keyfig0}(B) 
for an illustration). Competitive abilities of microbes may in general be 
different for different nutrients.

The final outcome of a competition of microbes for nutrients 
is a stable state in which no microbe can switch to a more 
preferred nutrient \emph{and simultaneously} win the competition 
with another microbe that is currently utilizing it. The microbial 
ecosystem will persist in this stable state until it is externally 
perturbed (e.g. by removal or addition of either microbes or nutrients).
Note that our definition of a stable state corresponds exactly to that 
in the original formulation of the stable marriage problem.

Inspired by the classical diauxic (or polyauxic) growth experiments 
\cite{Monod1949} we assume that microbes are constantly scouting 
the environment for more preferred nutrients. However, the diauxic 
shift down to the next nutrient requires the currently consumed (more preferred) 
nutrient to either be completely exhausted or at least to fall below 
a certain concentration threshold.  In what follows, we ignore the kinetics 
of this switching behavior including the lag phase. The natural microbial 
ecosystems relevant to our model may have rather complex dynamical behaviors 
including long transients, oscillations, and even chaos 
\cite{Balagadde2005,Zamamiri2001,Graham2007,Skupin2015}. However, 
these lie beyond the scope of the SMP-based approach.


Microbial preferences towards nutrients typically follow the order of maximal growth rates 
reached when they are present in a high concentration \cite{Aidelberg2014}. Using this as a general rule of thumb, we assume that a microbial species' stable-state abundance systematically decreases as it shifts down its nutrient preference list. The exact procedure by which we assign abundances to species in a stable state is described in Methods: Studying complementarity through different ranked interaction tables.





\subsection*{Community restructuring following external perturbations}
\begin{figure}
\includegraphics[width=\linewidth]{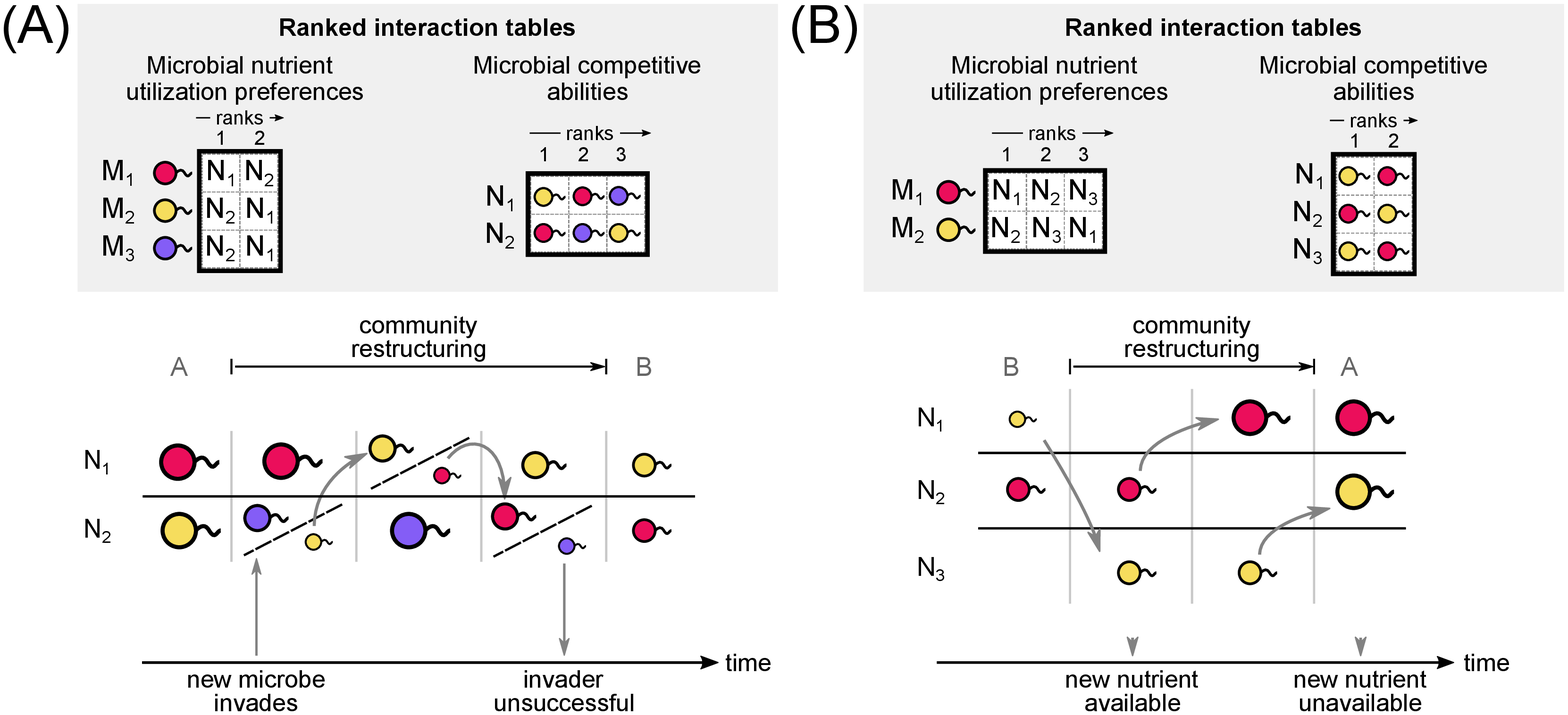}
\caption{\textbf{Community restructuring following external perturbations.}
\\ 
Two ranked tables of microbes' nutrient utilization preferences and competitive abilities 
are shown on top of each panel.
Colored circles represent different microbial species $M_1$, $M_2$, $M_3$. The size of each circle corresponds to the rank of a nutrients microbe currently utilizes - bigger sizes correspond to better ranks and thus larger populations. Different nutrients are labeled $N_1$, $N_2$, $N_3$. Oblique dashed lines indicate transient states for microbial competition. 
\textbf{(A)} The introduction of a new probiotic microbe, purple species ($M_3$), 
causes red ($M_1$) and yellow ($M_2$) species to enter into a competition with the invader. The dynamics of the stable marriage model results in a community restructuring to the state B, such that the red ($M_1$) and yellow ($M_2$) species shift their currently utilized nutrients to  their second choices. The invading purple species ($M_3$) fails to establish itself and disappears form the system \textbf{(B)} A transient addition of a prebiotic nutrient, $N_3$, restructures the community from state B back to state A, in which each microbe once again uses its most preferred nutrient. 
}
\label{keyfig1}
\end{figure}

We first consider a simple case in which two microbial species ($M_1$: red and $M_2$: yellow in figure \ref{keyfig1}) utilize two nutrients ($N_1$ and $N_2$). The preferences of microbes for these nutrients are complementary to each other: $M_1$ prefers $N_1$ to $N_2$, while $M_2$ prefers $N_2$ to $N_1$. The competitive abilities of microbes are opposite to their preferences. As shown in figure \ref{keyfig1} $M_2$ wins over $M_1$ in a competition for $N_1$, while $M_1$ wins over $M_2$ in a competition for $N_2$. There are two possible states of this ecosystem characterized by nutrients: the state A (see figure \ref{keyfig1}), where $M_1$ is consuming $N_1$ while $M_2$ is consuming $N_2$, and the state B, where $M_1$ is consuming $N_2$, while $M_2$ is consuming $N_1$. One can easily check that both states are stable in the SMP sense. That is to say, no microbe could switch to a nutrient it prefers more than the one it currently utilizes and simultaneously win the battle with another microbe which is its current consumer. The state A is the one obtained by the "microbe-proposing" algorithm. It naturally emerges whenever the current set of microbes is introduced to the system when all nutrients are 
supplied at a high influx. In this case, microbes following the sequence of diauxic shifts end up in this state and remain there until perturbed by addition of other microbes or nutrients, or (possibly transient) removal of the existing ones. The stable states in our model satisfy the criteria for alternative states of an ecosystem proposed in Ref. \cite{Connell1983}.

In what follows we investigate the stability of stable states in the example illustrated in figure \ref{keyfig1}, 
with respect to two types of perturbation: the introduction of a probiotic (another microbe $M_3$ shown in 
purple in figure \ref{keyfig1}(A)) and a prebiotic (a transient nutrient $N_3$ in figure \ref{keyfig1}(B)). 

In the case of the probiotic, the community starts at the state A - a natural endpoint of diauxic shifts. A new microbe $M_3$ (probiotic) is introduced to the community and initially displaces $M_2$ in the competition for its preferred nutrient, $N_2$. As a result, $M_2$ switches over to its next preferred nutrient ($N_1$) and outcompetes $M_1$, which was consuming it. $M_1$ now also switches to its second preferred nutrient $N_2$ and competitively displaces the `invader' $M_3$. $M_3$ switches to its second nutrient $N_1$ but loses the competition with $M_2$ and ultimately disappears from the system. Thus, in spite of its temporary success, the microbe $M_3$ fails to establish itself in the community. Note, however, that the result of its transient residence was a restructuring of the community from one stable state (A) to another (B). While the initial state A was `microbe-optimal' (i.e. both microbes consumed their most preferred nutrients in any of the stable states), the transient competitive interactions due to a new microbe pushed the community to a less microbe-optimal stable state, B. 

In the other illustrative case the community starts in the stable state B, driven there e.g. by consumption of a probiotic microbe (figure \ref{keyfig1}(A)). A new nutrient $N_3$ (prebiotic) is transiently added to the diet. The microbe $M_2$ prefers $N_3$ to its currently consumed nutrient ($N_1$) and switches to consume it. The $N_1$ is now available without competition, so microbe $M_1$ switches to use it as it stands higher than its currently consumed nutrient ($N_2$) in $M_1$'s preference hierarchy. 
After some time the prebiotic $N_3$ is removed from the diet. The microbe $M_1$ now switches to $N_2$ (its second preferred choice after $N_3$). Thus the community undergoes a restructuring again, this time from microbe-pessimal state B to microbe-optimal A. 

These examples illustrate the following general rule: the introduction of microbes and nutrients pushes the community structure in two opposite directions. Specifically, invading microbes increase competition for nutrients and generally result in a community restructuring towards a stable state that is less growth-optimal for microbes. Even short-lived introduction of extra nutrients, on the other hand, relieves this competition and restores the community towards 
stable states in which microbes use more preferred nutrients.

\subsection*{\textcolor{black}{Multiple stable states and the network of transitions between them}}

\begin{figure}
\includegraphics[width=\linewidth]{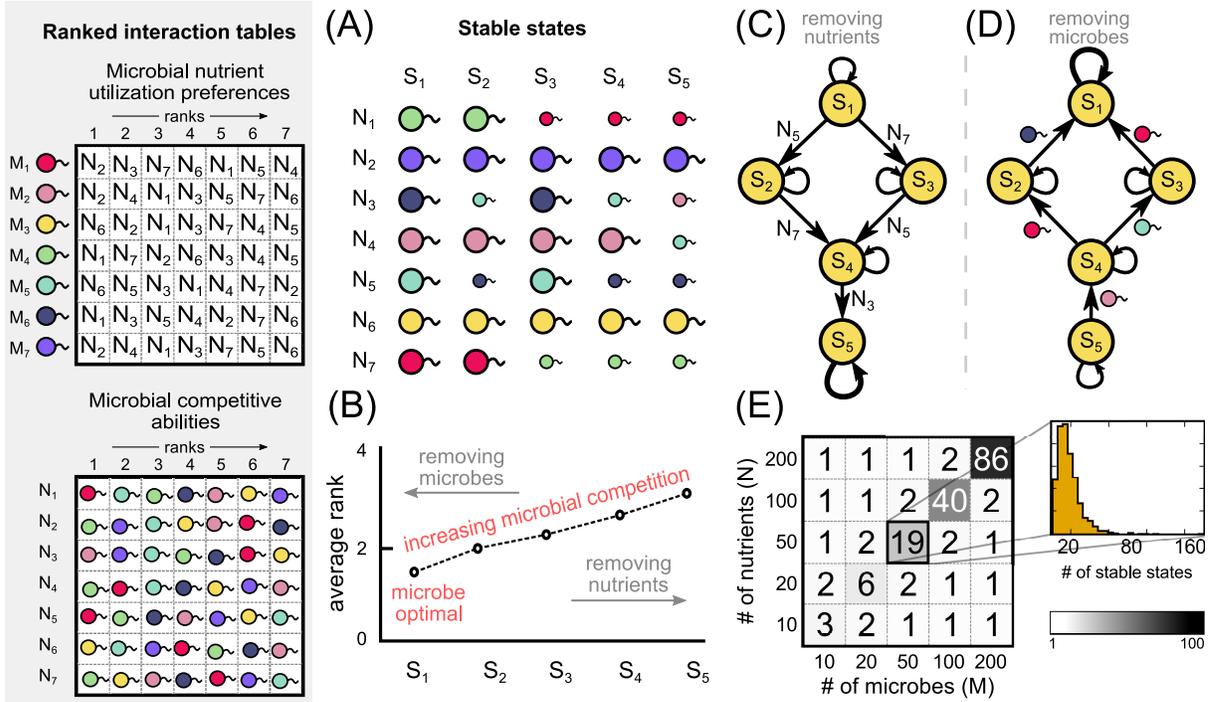}
\caption{\textbf{\textcolor{black}{Multiple stable states and the network of transitions between them.}} 
Two ranked tables of microbes' nutrient utilization preferences and competitive abilities 
are shown on the left. 
\textbf{(A)} 
The list of all stable states (labeled $S_1$ through $S_5$) in the model. 
In each stable state, every microbe (colored circles with tails; sizes indicative of how preferred the consumed nutrient in a state is) exclusively consumes one nutrient (labeled $N_1$ through $N_7$). 
\textbf{(B)} The 'microbe-optimality' of stable states $S_1-S_5$ (lower is better for microbes) 
quantified by the rank of the consumed nutrient averaged over all microbes.
Microbe-optimality can be improved by transiently removing microbes and deteriorated by transiently removing nutrients.
\textbf{(C, D)} The stable states are 
connected via the `restructuring network' of transitions. The community in the model gradually restructures from $S_1$ towards $S_5$ by transient nutrient removal (for details, see Results: Higher-order interactions enable multiple stable states) and from $S_5$ back towards $S_1$ by transient microbe removal. In this restructuring network, a pair of stable states is connected by a directed link, if the community can transition between these states via transient removal of 
just one nutrient (removed nutrient and directionality are shown in panel (C)) or of a single microbe (removed microbe and directionality are shown in panel (D)). 
\textbf{(E)} Average number of stable states for communities with different numbers of microbes ($M$, x-axis) and nutrients ($N$, y-axis) and randomized interaction tables. (Inset, top) For $(M, N)$ = $(50, 50)$, we show that the number of steady states (in orange) for 1,000 random interaction tables. 
\textcolor{black}{(Inset, bottom) The number of steady states as a function of $M$ (when $M=N$).}
}
\label{keyfig2}
\end{figure}



In general, the number of stable states increases with the number of microbes 
and nutrients in the community. In figure \ref{keyfig2} we show an example of 
a community where 7 microbial species compete for 7 distinct nutrients, all of 
which they can utilize. For \textcolor{black}{the} particular set of microbial 
nutrient preferences and competitive abilities shown 
as ranked tables in figure \ref{keyfig2}(A), 
there are a total of 5 stable states labeled $S_1$ through $S_5$.

As understood in the context of the original stable marriage problem 
\cite{Gusfield1989}, the stable states can be arranged hierarchically 
in the order of decreasing microbe-optimality 
quantified by the average rank of nutrients consumed by microbes in a particular state.
Since rank $1$ corresponds to the most preferred nutrient, while rank $N$ 
corresponds to the least preferred one, lower values for this optimality 
measure correspond to more microbe-preferred states. 
%
The labels of the states $S_1-S_5$ were arranged in the order of decreasing microbe-optimality , i.e. increasing the average rank of consumed nutrients (see figure \ref{keyfig2}(B)). Thus the state $S_1$ is the most optimal for microbes (corresponding to the stable state generated by the 'microbe-proposing' Gale-Shapley algorithm in the SMP), while the state $S_5$ 
is the least optimal one. The average rank of consumed nutrients in $S_1$ is equal to 1.7 which means that even in this state, not every microbe gets its most preferred nutrient.
This should be compared to its value $\sim 2.9$ in the state
$S_5$, where a typical microbe gets its third choice among nutrients. 

As described in Methods, the transitions between stable states of the SMP can be realized by transiently breaking a `marriage', i.e. disrupting a microbe-nutrient pair. Figure \ref{keyfig2}(B) shows that 
the removal of nutrients from a diet (starvation) generally drives the community further 
away from the microbe-optimal state ($S_1$). Indeed, in this case (akin to the probiotic 
case shown in figure \ref{keyfig1}(A)) microbes need to compete more for the remaining 
nutrients. Removing a specific subset of microbes (e.g. by antibiotics) has the opposite 
result: the surviving microbes have fewer competitive interactions for nutrients and hence 
each one of them would get a better (or same) ranked nutrient according to its preference list. 
\textcolor{black}{Thus, somewhat counterintuitively, transient introduction of antibiotics might shift a community 
towards a more microbe-optimal state with larger overall biomass,  
which can be experimentally verified.} 

As known from the SMP results,  the transitions between stable states could be triggered only 
by the removal of a very specific subset of nutrients or microbes. These states can thus be 
arranged in a `community restructuring network' shown in figure \ref{keyfig2}(C, D).  The 
transition along a given edge of this network leading further away from the microbe-optimal 
state could be triggered by a transient removal of a specific single nutrient (figure \ref{keyfig2}(C)). 
The transition in the opposite direction (towards a microbe-optimal state) is triggered by the 
transient removal of a specific single microbial species (see figure \ref{keyfig2}(D)). 
Removal of a nutrient leaves the microbe that was utilizing it temporarily without its 
source of energy. This microbe will then engage in competition with other microbes for the 
remaining nutrients. This results in a cascade of shifts where microbes begin to utilize 
less-preferred nutrients, as prescribed by the Gale-Shapley algorithm. If the removed 
nutrient is  reintroduced soon after its removal, the community will return back to its 
original state, contributing to the community's resilience. In the opposite case, if the 
nutrient's absence lasts very long, one of the microbial species left without a nutrient will 
go extinct. However, there is a specific intermediate regime where the nutrient is reintroduced 
at \emph{just} the right time for its microbial consumer in the new stable state to have 
recently switched towards it. In this case, such a transient nutrient removal results 
in a community restructuring from a stable state to another one but less microbe-optimal. 
A similar restructuring is possible when a microbial species is transiently removed from 
the community (e.g. by a narrow-spectrum antibiotic) so the nutrient it utilized before 
the removal is now open for competition from other microbes. If this microbe is reintroduced 
later at just the right time, the community can restructure towards another stable state 
which is more microbe-optimal.

These examples (as well as their counterparts in which microbes or nutrients were added 
to the community as discussed in the previous section and illustrated in figure \ref{keyfig1}) 
demonstrate that these stable states are relatively resilient with respect to many transient 
perturbations. Such resilience is exhibited at two different levels. Firstly, not all 
perturbations result in community restructuring. Those perturbations that \emph{do} arrange 
the stable states in a hierarchical `community restructuring network' \textcolor{black}{are} 
shown in figure \ref{keyfig2}(C). For any two adjacent stable states in this network, there is 
just one specific nutrient and one specific microbe that can be removed to trigger a transition 
between them. Transient removal of other nutrients or microbes is shown as self-loops in figure 
\ref{keyfig2}(C, D), since these events return the community back to the original stable state.  
Secondly, even when this carefully selected nutrient or microbe is removed, it must be reintroduced 
within a specific time interval (not too soon and not too late) to result in a successful restructuring.

The average number of stable states for different combinations of numbers of microbes, $M$, 
and nutrients, $N$, are shown as a grid in figure \ref{keyfig2}(E). { \color{black}The distribution of the number of stable states for different (random) realizations of microbe preference lists and competitive abilities for $M=N=50$ is shown in the orange histogram in the inset to figure \ref{keyfig2}(E). Further, supplementary figure S3 shows: (1) how the number of stable states decreases when the correlation among preference lists increases (figure S3(C)--(E)); and (2) how the average number of stable states increases with increasing $M$ or $N$ (figure S3(A)).
}

\subsection*{Complementary prioritization of nutrients as a mechanism for robust many-species coexistence}
\begin{figure}
\includegraphics[width=\linewidth]{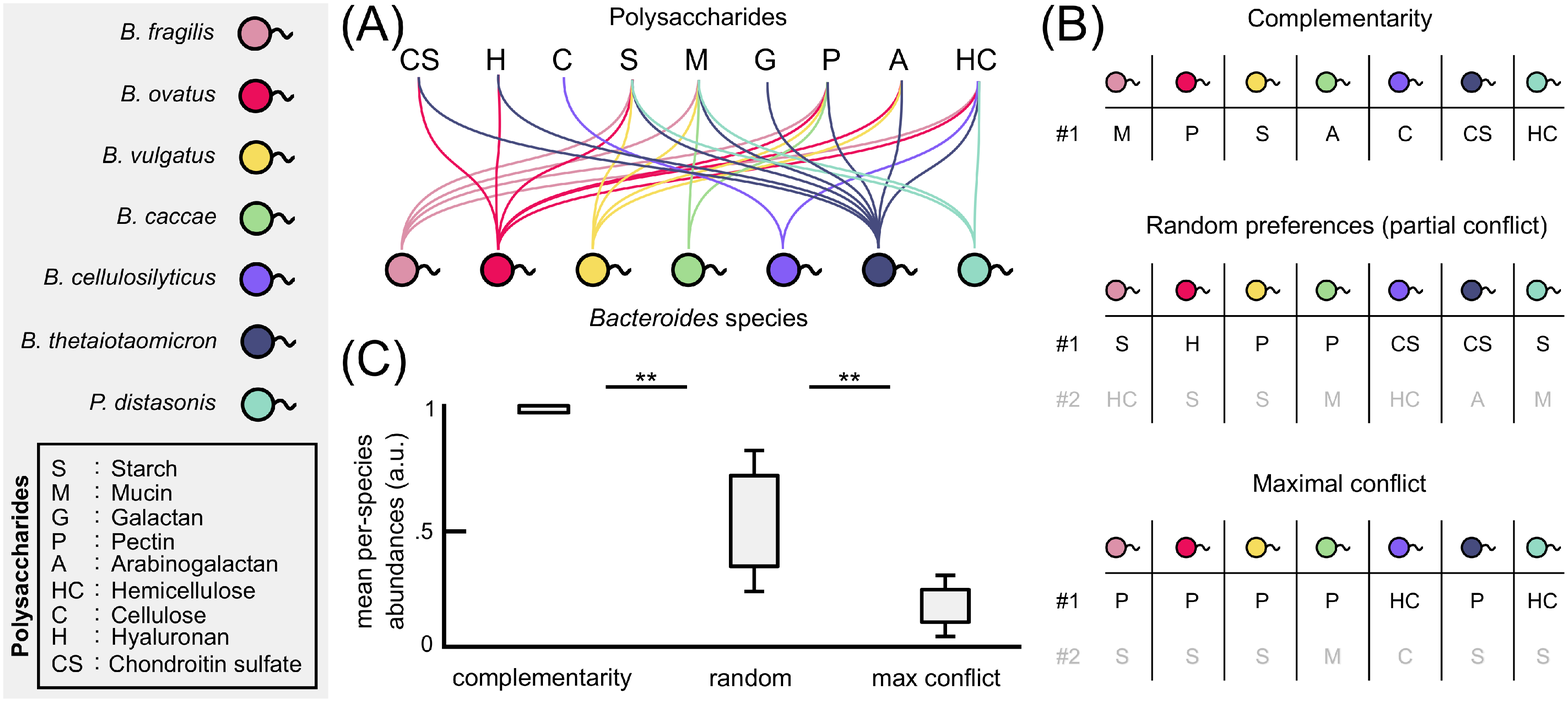}
\caption{\textbf{Complementary polysaccharide prioritization allows robust coexistence in gut \emph{Bacteroides} species.} \\
\textbf{(A)} The polysaccharide utilization network of \emph{Bacteroides} species in the human gut (data taken from\cite{Chia2017}). The character labels represent 9 different polysaccharides: starch (S), mucin (M), galactan (G), pectin (P), arabinogalactan (A), hemicellulose (HC), cellulose (C), hyaluronan (H), chondroitin sulfate (CS) --- known to be frequently present in human diets (legend in the box on the left), whereas the colored circles represent 7 different \emph{Bacteroides} species 
routinely found in human gut microbiome: \emph{Bacteriodes fragilis}, \emph{B. ovatus}, \emph{B.vulgatus}, \emph{B.caccae}, \emph{B.cellulosilyticus}, \emph{B. thetaiotaomicron}, \emph{Parabacteroides distasonis}.
Undirected links between microbes and polysaccharides indicate a species' ability to metabolize that polysaccharide. 
\textbf{(B)} 
Examples of microbial nutrient preferences (the most preferred nutrient of each of the microbes) are sorted into three categories: complementary (top) where microbes' top preferred nutrients 
(\#1) 
are all distinct from each other; random (middle) preferences where all ranked lists are randomly generated;
and maximal conflict (bottom) which represents the maximum intersection between the sets of top 
(\#1) 
and second 
(\#2) 
preferred nutrients of different microbes.
\textbf{(C)} For 1,000 randomly sampled microbial preferences from each category, we simulated the stable marriage model to compute the expected per species microbial abundances (see Methods: Studying complementarity through different ranked interaction tables) for each case as box plots. The box plots quantify the distribution of average microbial abundance assumed to be inversely proportional to the rank of utilized nutrient. The average abundance is the largest in the case of complementary nutrient choices,  
All differences between distributions of abundances in each category
are highly statistically significant according to 
the Kolmogorov-Smirnov test 
with a $P$-value threshold of $0.01$.
}
\label{keyfig3}
\end{figure}

The human gut microbiome provides a fertile testing ground for 
our model. Indeed, as discussed in the introduction, many gut microbes are known to 
utilize nutrients sequentially.
Moreover, recent reports indicate that multiple \textit{Bacteroides} species have been 
regularly observed at high abundances simultaneously, in spite of a strong  overlap in 
their metabolic capabilities \cite{Raghavan2015}. This overlap is visualized in figure 
\ref{keyfig3}(A), where we show a network connecting each of 7 
abundant species 
in the human gut (\emph{Bacteriodes fragilis}, \emph{B. ovatus}, \emph{B.vulgatus}, 
\emph{B.caccae}, \emph{B.cellulosilyticus}, \emph{B. thetaiotaomicron}, and a 
recently reclassified member of 
the \emph{Bacteroidetes} phylum \emph{Parabacteroides distasonis})
with a subset of 9 polysaccharides (starch, mucin, galactan, pectin, arabinogalactan, 
hemicellulose, cellulose, hyaluronan, chondroitin sulfate) they are capable of utilizing 
as energy sources (data from\cite{Chia2017}, see Methods for details). 
For \textcolor{black}{the sake of brevity,} in what follows we refer to this set as \textit{Bacteroides} species.
What strategies by these microbes would allow their `robust' co-occurrence in the human gut, 
i.e. long-term, stable coexistence at high abundances?

The stable marriage problem provides a natural framework in which to look for such strategies. 
Indeed, by supplementing the utilization network shown in figure \ref{keyfig3}(A) with a 
specific set of ranked nutrient preferences and competitive abilities of all participating 
microbial species, our model can predict which species will survive, how many stable states 
the corresponding community can be in, and what kind of abundance profiles they will achieve 
in these states. The latter could be approximated by the inverse of the rank of the consumed 
nutrient for every surviving microbe in a particular stable state. Indeed, microbes utilizing 
their preferred (low rank) nutrient are expected to reach high abundances. It stands to reason 
that in order to simultaneously achieve high abundances, these species have to successfully 
partition the set of nutrients among themselves. In the presence of a strong metabolic overlap 
this requires microbes to have evolved a mutually complementary set 
of nutrient preferences. 

We quantify the complementarity of microbes' top preferences by calculating 
the number of competing pairs 
of microbes that have the same most-preferred nutrient. This number can vary between 0 
(for perfect complementarity; figure \ref{keyfig3}(B) [top case]), to around 6 
(for random preferences; figure \ref{keyfig3}(B) [middle case]) and 
ultimately up to 11
(for maximal conflict in these lists; figure \ref{keyfig3}(B) [bottom case]). 
\textcolor{black}{The maximal conflict case 
assumes the strongest possible similarity of the entire preference lists 
of different microbes (see Methods for details).} 
%
%

\textcolor{black}{We tested 1000 preference lists from each of these 3 categories (complementary, random 
and maximal conflict) and calculated the average microbial abundances in each case 
(see box plots in figure \ref{keyfig3}(C)).} As expected, the average abundance is 
the highest in the case of complementarity, lower for random preferences, and 
lower still for maximal conflict. {\color{black}Moreover, communities with complementary preferences show a higher number of stable states (see figure S2(B) for \emph{Bacteroides} and figure S3(C)--(E) for a more general result for the communities in our model).}

Perfect complementary between the top preferences of 7 microbes would require careful 
orchestration over evolutionary times. However, these choices are encoded in regulation 
of specific Polysaccharides Utilization Loci (PULs) controlled by microbial transcription 
regulatory networks and have been shown to be quite flexible \cite{Raghavan2015}. Thus 
the complementarity of top nutrients choices required for robust coexistence of 
\textit{Bacteroides} species in the human gut is entirely plausible and, indeed, has been 
in part reported in Ref. \cite{Martens2017}.


\section*{Discussion}

In the results presented above, we describe a conceptual model of microbial competition 
for sequentially utilized nutrients. This model can exhibit rich behaviors such as 
dynamic restructuring and multiple stable states connected by a hierarchical 
transition network. All of this complexity is encoded in just two ranked tables: 
one with microbial nutrient preferences and the other with their competitive abilities 
for different nutrients.  The competitive interactions summarized in these tables are 
just starting to be explored experimentally. In fact, the first experimental results 
relevant to communities within the human gut have already been reported \cite{Martens2017, Riviere2018}. 
Specifically, these results demonstrate the preferences and competitive abilities of 
2 \textit{Bacteroides} species for 9 particular polysaccharides.

In the absence of experimentally determined preferences, the \textcolor{black}{naive} 
expectation would be to use randomized nutrient preferences and competitive abilities. 
However, as shown in figure \ref{keyfig3}, the results for random preference tables 
qualitatively disagree with experimental observations of robust coexistence of multiple 
species (e.g. \textit{Bacteroides} in human gut) competing for the same set of nutrients. 
Our model shows that complementarity in nutrient preferences of 
\textcolor{black}{different microbes facilitates} such coexistence. This is consistent with 
\textcolor{black}{experimental studies reporting that frequently co-occurring microbial species 
tend to have complementary nutrient preferences} \cite{Comstock2015, Riviere2018, Martens2017}. 

\textcolor{black}{Complementary nutrient preferences may} also explain the prevalence 
of habitat filtering in many naturally-occurring microbial communities 
\cite{Weiher1998, Cornwell2006, Levy2013, Goldford2017}, i.e. the observation that 
many metabolically overlapping species stably coexist with each other. This apparently 
paradoxical observation is unsurprising in the light of our results
\textcolor{black}{assuming that nutrient preferences of these species co-evolved to be 
(at least partially) complementary to each other.}

\textcolor{black}{One factor complicating the (co-)evolution of nutrient complementarity 
is that certain nutrients tend to be universally prized by all microbes.
This is true for simple, easy-to-digest metabolites with high-energy content 
(e.g. simple sugars) where evidence suggests the existence of a common 
preference order \cite{Ammar2018}.
However, the order of microbial preferences for more complex, harder-to-digest 
nutrients such as polysaccharides is known to be much more flexible 
\cite{Martens2017, Riviere2018}. 
}

{
\color{black}
Correlations or complementarity in the preference lists of different microbial species 
(correlations of Type A) discussed above are just one 
out of three types of correlations possible in our model. The remaining two 
correlations are: Type B --- Correlations in competitive abilities of different microbes; 
and Type C --- Correlations between each microbe's nutrient preferences and 
competitive abilities for the same nutrient.
Strong positive correlations of type B imply the existence of ``super bugs'' 
good at utilizing every resource. Conversely, negative type B correlations may 
arise due to tradeoffs in each microbe's competitive abilities for 
different nutrients \cite{Wingreen2017}. For type C only the positive correlations 
are biologically plausible. Indeed, one might expect microbes to have higher-than-average 
competitive abilities for those nutrients that they prefer to consume first. 

Positive correlations of all three types reduce the number of stable states, ultimately resulting in a unique stable state for fully correlated lists (see figure S3, panels (C) to (E) for correlations of types A, B and C respectively). Supplementary Fig. S3 explores the model with correlated lists by 
plotting the number of stable states as a function of correlation strength. 
}

\textcolor{black}{
An evolutionary variant of the stable marriage model allows one to answer 
questions related to metabolic specialization of microbes. These questions include:
How many nutrients a given microbe should have the capacity of using? That is to say, how 
many distinct nutrient utilizing metabolic pathways should be encoded in its genome?
How do different microbes make a choice between being broad generalists and narrow specialists? 
In our analysis we see examples of both among \textit{Bacteroides} species (see figure \ref{keyfig3}(A)). 
The common wisdom is that in stable environments, characterized by a reliable influx 
of the same set of nutrients, microbes tend to become narrow specialists. 
However, this strategy would not fare well for microbes trying to survive 
in strongly fluctuating environments, where each microbe needs to be able 
to switch between multiple nutrients until it finds one 
currently present in the environment.
An intriguing possibility is that the evolutionary trajectory 
of each species may be shaped by its partners in the stable marriage problem. 
That is to say, given its microbial partners, there is no need for 
a microbe to retain metabolic pathways utilizing nutrients which it never gets to use 
in any of the ``stable marriages''. Over evolutionary time, such unused pathways would 
be dropped from its genome. At the same time, microbes would tend to improve their 
competitive abilities for the remaining nutrients, which in turn could possibly 
reinforce the initial set of stable states in the ecosystem. 
Hence the stable states in the marriage model may leave their footprint on 
the genomic content of co-evolved microbial species. More technically, in this case 
the set of nutrients each microbe could utilize would coincide with its 
``reduced Gale-Shapley preference list'' (see \cite{Gusfield1989} for definitions).
}


\textcolor{black}{Our model assumes one-at-a-time sequential consumption of all nutrients by all microbes. 
However, real-life microbes are known to combine sequential consumption and 
co-utilization of different nutrients depending on the topology of their 
catabolic pathways \cite{Wang2017}.
We can potentially incorporate 
co-utilization of nutrients to our model as 
a ``many-to-many'' matching rules \cite{Gusfield1989} 
combined with ties in the ranked lists 
\cite{Miyazaki2015}. The number and nature of 
stable states in such models 
remain to be explored in a future study.}   

\textcolor{black}{Furthermore,} a key driver of diversity in real-life microbial communities often lies in the metabolic byproducts generated by resident species. Indeed, in the presence of metabolic byproducts the number of microbial species in the steady state is no longer limited from above by the number of externally provided nutrients. Recent models \cite{Goyal2017, Goldford2017} and experiments \cite{Antonopoulos2017, Goldford2017} demonstrate that a diverse microbial ecosystem may be supported even by a single externally provided nutrient.  The \textit{Bacteroides} species used in our study are also known to grow on each other's metabolic byproducts \cite{Comstock2015}. That may be the reason why \textit{B. thetaiotaomicron} survives while losing the competition to \textit{B. ovatus} on all 8 polysaccharides studied in reference \cite{Martens2017} (see figure 4 from that reference). 

\textcolor{black}{
The basic stable marriage model allows for a natural multi-layered 
generalization involving cross-feeding between microbial species. 
One starts with a single layer composed of abundant primary nutrients, 
which for human gut include polysaccharides shown in figure \ref{keyfig3}(A).
The microbes (such as \textit{Bacteroides} species in figure \ref{keyfig3}(A)) 
compete, or, alternatively, complementarily utilize these nutrients and generate the second layer of nutrients, composed of their metabolic byproducts (or products of extracellular metabolic degradation). These byproducts in turn allow for a new set of microbes to grow and generate yet another layer of byproducts. 
Furthermore, microbes from the upper layers would normally not 
compete for nutrients in the layers below them. Indeed, the concentration of nutrients is expected to rapidly decrease with a trophic layer \cite{Goyal2017}. Hence, to maximize their growth rate, microbes would prefer nutrients from higher trophic layers.
} 

\textcolor{black}{
Microbes using nutrients one-at-a-time give rise to 
tree-like food webs similar to those studied in Ref. \cite{Goyal2017}. 
In our case there will be multiple trees, each growing from a 
single primary nutrient. These trees would generally change as the community switches 
from one stable state to another. All the results of Ref. \cite{Goyal2017} 
including the functional forms of the distributions of species' 
abundances and prevalences 
are directly transferable to the multi-layered variant 
of the stable marriage model. 
}

\textcolor{black}{
Another generalization of our model is when 
nutrients come in two or more distinct types, each essential for 
microbial growth (e.g. carbon and nitrogen sources). 
An extension of the model in this case would require 
a microbe to choose one source of each type. This would 
correspond to a marriage with more than two sexes. 
As far as we know, these modifications of the 
stable marriage model have not been developed yet, 
though this possibility has been explored in works of 
science fiction \cite{Asimov1990Gods,Butler1989xenogenesis}.
}


\textcolor{black}{A natural way to think about the competition in the stable marriage context is in terms of species and nutrients subject to a constant dilution in a chemostat. Changing the dilution rate would drive the ecosystem through different qualitative regimes of nutrient utilization. Another possible realization is in a periodically diluted batch culture where the system is diluted and the nutrients are added at discrete time points 
in a cyclic fashion.
When thinking about such batch-fed bioreactors, one needs to consider 
the possibility of transient co-utilization of the same nutrient 
by several microbes. How can our model adapt to this possibility? 
One of the variants of the stable marriage problem known as the 
hospitals/residents problem \cite{Gale1962, Gusfield1989, Roth1992} 
provides a possible starting point for such adaptation. In this 
problem a hospital (a nutrient in our case) can accommodate multiple 
residents (microbes). A variant of the Gale-Shapley algorithm 
\cite{Irving2000, Manlove2002} allows one to find all stable states 
of the community. Most other mathematical results of a ``pure'' marriage 
problem are also directly transferable here with only minor modifications. 
}


\textcolor{black}{
Another appealing feature of our model is that it naturally incorporates higher-order interactions between microbial species \cite{Billick1994}. These interactions
have recently brought to attention after a large number of studies showed 
that pairwise interactions are not sufficient to explain community dynamics \cite{Momeni2017, Friedman2017, Goldford2017, Venturelli2017}. Further, they have been implicated as an important factor contributing to the composition, stability, and diversity of ecosystems \cite{Levine2017, Grilli2017, Bairey2016, Golubski2016}. 
%
In our model, community dynamics depend on ranked preferences and competitive abilities of all species in a resource-explicit manner, and cannot be simply reduced to a set of pairwise competitive outcomes.
%
That is to say, the outcome of the competition between species can be rather different depending on the presence or absence of other species. This is reflected in different species abundance profiles in figure \ref{keyfig2}(A) (see for instance, states $S_2$ and $S_3$).
}

\textcolor{black}{To summarize, in this study we present a model inspired by the stable marriage problem that shows and gives insights regarding several dynamic microbial community phenomena. These phenomena include the observation of several stable states, dynamics of transitions between these states, as well as how they restructure.
The stable states in our model satisfy all three necessary criteria for alternative stable states set forth in Ref. \cite{Connell1983}. 
Our model assumes that several microbes utilize nutrients sequentially (diauxie or polyauxie). With this assumption, all the stable states of a community are fully determined by two ranked tables: one summarizing all microbes' preferred order of utilization of nutrients, and the other their competitive ability to uptake these nutrients relative to others microbes. 
Such rank tables can be inferred from polyauxic shift experiments in which individual microbes are grown on a rich medium with many nutrients. Further experiments in this direction will help generate predictions against which to test our model.}

\section*{Methods}
\subsection*{Enumerating all stable states}
For any general case of preference lists in the stable marriage problem (SMP), there exist multiple `stable states'. There are several algorithms to enumerate all these states, though we used in our study one that is intuitive and connects well with microbial communities: the so-called `breakmarriage' algorithm \cite{McVitie1971, Gusfield1989}. 
For our problem this algorithm involves 
starting from one of the stable states (e.g. microbe-optimal one) and then 
successively \emph{breaking} each of the microbe-nutrient pairs by 
removing either a microbe or a nutrient. A transient removal of a 
specific nutrient has the possibility of triggering a transition of the 
community to another stable state in which all microbes are worse off (or equal) in terms of the preference rank of the nutrient they consume. These transitions are shown as 
downward pointing arrows in Fig. \ref{keyfig2}(C). Conversely, a transient removal of a 
specific microbe  could trigger a transition to a stable state in which all microbes are better off (or equal) in terms of the preference rank of the nutrient they consume (upward pointing arrows in Fig. \ref{keyfig2}(D)). Below, we list the specific details of the 
`breakmarriage' algorithm.



One starts with the microbe-optimal stable state obtained through the Gale-Shapley algorithm (see \cite{Gale1962}) in which every microbe plays the role of the active party and thus gets the best nutrient in any stable state. In the example illustrated in figure \ref{keyfig2}(B), this corresponds to the state $S_1$. One chooses an implicit ordering of microbes (say for convenience, in increasing order from $M_1$ to $M_M$ for $M$ microbes) in which one attempts to break microbe-nutrient pairs. 

Upon breaking a pair (in our example, $N_5$ and the teal microbe $M_5$), the microbe in that pair ($M_5$) is left without a nutrient, and therefore shifts down to (i.e. `proposes marriage to' in the SMP jargon) the next nutrient in its preference list ($N_3$). If $M_5$ is more competitive than the current consumer of this nutrient (the dark blue microbe, $M_6$) with respect to the nutrient $N_3$, it competitively displaces this current consumer ($M_6$). (If not, the microbe ($M_5$) continues to shift down its preference hierarchy until it finds a nutrient it can utilize.) Every time a microbe is left without a nutrient, it continues to down-shift its nutrient preference list and attempts to competitively displace other microbes using these nutrients (in our example, $M_6$ now moves to attempt to use $N_5$). If along this sequence, the original nutrient whose pair was broken ($N_5$) is `proposed' to by another microbe (here, by $M_6$), and if $M_6$ {\it can competitively displace  its original partner} ($M_5$ in our case), a `rotation' is said to have been successfully completed and the new state is guaranteed to be stable (here, that state is $S_2$ shown in Fig. \ref{keyfig2}(B)). If any of these steps fails, the attempted rotation is unsuccessful and one reverts back to the previous stable state and then attempts to break the \emph{next} microbe-nutrient pair according to our implicitly chosen order.

For any of the new stable states (say $S_2$ described above) found through this procedure, one repeats this procedure using this state as the initial stable state to find even more stable states. When all microbe-nutrient pairs in all such obtained stable states have been attempted to be broken, the algorithm is terminated. This procedure is guaranteed to enumerate all possible states for a chosen set of ranked interaction tables.

\subsection*{Studying complementarity through different ranked interaction tables}
We sampled a large number of possible interaction tables, i.e. preferences towards nutrients and competitive abilities for all gut microbes of the genus \emph{Bacteroides} regularly found at high abundances in the human gut (data taken from \cite{Chia2017}). 

In principle, there are close to $10^{131}$ such possibilities, and it is thus not possible to sample \emph{all} such tables. Instead, we compartmentalize such interactions in three broad categories: complementary, random and maximal conflict.

In complementary interaction tables (see figure \ref{keyfig3}(C) [top case]), we construct random interaction tables with the following constraint: microbial preferences for the top (most preferred) nutrient must be made maximally distinct, i.e. with no overlap if possible. To construct interaction tables in this category, we begin by picking a microbe at random and assigning it a nutrient it can utilize at random. We then remove this nutrient as a possible top choice for all other microbes. We then randomly pick another microbe (without replacement) from the full set and assign it another random nutrient. We continue this until all microbes have been assigned a distinct most preferred nutrient. In case a chosen microbe has no choice left, we discard that particular interaction scenario and start a new one.

Random interaction tables provide a null interaction scenario for our model (see figure \ref{keyfig3}(C) [middle case]) and are thus used to set the na\"ive expectation for competition and conflict within these gut microbes. In this scenario microbial preferences towards nutrients are selected by a random permutation independently chosen for each of the microbes.  

In interaction tables with maximal conflict (see figure \ref{keyfig3}(C) [bottom case]), we construct random interaction tables with the following constraint: we attempt to maximize the number of conflicting pairs (NCP) for the set of microbes (see Results: Complementary prioritization as a mechanism for robust many-species coexistence). For this, we pick a microbe at random and then randomly pick a nutrient it can utilize as its most preferred (top choice). For all other microbes in our set that can utilize this nutrient, we set it as their most preferred nutrient as well. We continue until all microbes have been assigned a most preferred nutrient and then randomize the rest of the interaction tables.

In all three cases described above the competitive abilities of microbes for each of the nutrients are set by a random permutation. 

Each specific pair of interaction rank tables (one for microbial preferences and another, for their competitive abilities) represents a possible competitive scenario in the human gut. We construct $1,000$ tables for each case. We then use the Gale-Shapley algorithm \cite{Gale1962} to find the microbe-optimal stable state of the possible \emph{Bacteroides} community and the breakmarriage algorithm (see Methods: Enumerating all stable states) to find the overall number of stable states. In the microbe-optimal state, we compute the relative rank of each microbe's utilized nutrient in their preference lists, i.e. the rank of the utilized nutrient relative to how many nutrients that microbial species is known to utilize. The inverse of this relative rank is used (in a.u.: arbitrary units) as a predictive measure of its species abundance in the resultant community. We repeat this procedure for all microbes in the community and then normalize the abundances of all microbes to add up to one so that the relative abundance for each species is between $0$ and $1$.

\section*{Acknowledgments}
A.G. acknowledges support from the Simons Foundation as well as the Infosys Foundation.
We thank Chen Liao for useful discussions of diauxic shift models. \textcolor{black}{We are also grateful to two anonymous reviewers whose comments and suggestions helped improve and clarify the manuscript.}

\section*{Author contributions}
S. M. conceived the study, A. G. and V. D. conducted the data analysis and modeling, A.G., V.D. and S.M. wrote the manuscript.

\section*{Competing financial interests} 
The authors declare no competing financial interests. 

\bibliographystyle{ismeapa}
\bibliography{arxiv_template}

\begin{thebibliography}{}

\bibitem[Advani et~al., 2017]{Advani2017}
Advani, M., Bunin, G., and Mehta, P. (2017).
\newblock Environmental engineering is an emergent feature of diverse
  ecosystems and drives community structure.
\newblock {\em arXiv preprint arXiv:1707.03957}.

\bibitem[Aidelberg et~al., 2014]{Aidelberg2014}
Aidelberg, G., Towbin, B.~D., Rothschild, D., Dekel, E., Bren, A., and Alon, U.
  (2014).
\newblock Hierarchy of non-glucose sugars in escherichia coli.
\newblock {\em BMC systems biology}, 8(1):133.

\bibitem[Ammar et~al., 2018]{Ammar2018}
Ammar, E.~M., Wang, X., and Rao, C.~V. (2018).
\newblock Regulation of metabolism in escherichia coli during growth on
  mixtures of the non-glucose sugars: arabinose, lactose, and xylose.
\newblock {\em Scientific reports}, 8(1):609.

\bibitem[Asimov, 1972]{Asimov1990Gods}
Asimov, I. (1972).
\newblock {\em The Gods Themselves}.
\newblock Doubleday.

\bibitem[Bairey et~al., 2016]{Bairey2016}
Bairey, E., Kelsic, E.~D., and Kishony, R. (2016).
\newblock High-order species interactions shape ecosystem diversity.
\newblock {\em Nature communications}, 7.

\bibitem[Balagadd{\'e} et~al., 2005]{Balagadde2005}
Balagadd{\'e}, F.~K., You, L., Hansen, C.~L., Arnold, F.~H., and Quake, S.~R.
  (2005).
\newblock Long-term monitoring of bacteria undergoing programmed population
  control in a microchemostat.
\newblock {\em Science}, 309(5731):137--140.

\bibitem[Billick e Case, 1994]{Billick1994}
Billick, I. and Case, T.~J. (1994).
\newblock Higher order interactions in ecological communities: what are they
  and how can they be detected?
\newblock {\em Ecology}, 75(6):1529--1543.

\bibitem[Briones e Raskin, 2003]{Briones2003}
Briones, A. and Raskin, L. (2003).
\newblock Diversity and dynamics of microbial communities in engineered
  environments and their implications for process stability.
\newblock {\em Current Opinion in Biotechnology}, 14(3):270--276.

\bibitem[Butler, 1989]{Butler1989xenogenesis}
Butler, O.~E. (1989).
\newblock {\em Xenogenesis Trilogy}.
\newblock Warner Books.

\bibitem[Caldarelli et~al., 2001]{Caldarelli2000}
Caldarelli, G., Capocci, A., and Laureti, P. (2001).
\newblock Sex-oriented stable matchings of the marriage problem with correlated
  and incomplete information.
\newblock {\em Physica A: Statistical Mechanics and its Applications},
  299(1):268--272.

\bibitem[Connell e Sousa, 1983]{Connell1983}
Connell, J.~H. and Sousa, W.~P. (1983).
\newblock On the evidence needed to judge ecological stability or persistence.
\newblock {\em The American Naturalist}, 121(6):789--824.

\bibitem[Consortium et~al., 2012]{HMP2012}
Consortium, H. M.~P. et~al. (2012).
\newblock Structure, function and diversity of the healthy human microbiome.
\newblock {\em Nature}, 486(7402):207--214.

\bibitem[Cornwell et~al., 2006]{Cornwell2006}
Cornwell, W.~K., Schwilk, D.~W., and Ackerly, D.~D. (2006).
\newblock A trait-based test for habitat filtering: Convex hull volume.
\newblock {\em Ecology}, 87(6):1465--1471.

\bibitem[Costello et~al., 2012]{Costello2012}
Costello, E.~K., Stagaman, K., Dethlefsen, L., Bohannan, B.~J., and Relman,
  D.~A. (2012).
\newblock The application of ecological theory toward an understanding of the
  human microbiome.
\newblock {\em Science}, 336(6086):1255--1262.

\bibitem[Coyte et~al., 2015]{Coyte2015}
Coyte, K.~Z., Schluter, J., and Foster, K.~R. (2015).
\newblock The ecology of the microbiome: networks, competition, and stability.
\newblock {\em Science}, 350(6261):663--666.

\bibitem[David et~al., 2014]{David2014}
David, L.~A., Maurice, C.~F., Carmody, R.~N., Gootenberg, D.~B., Button, J.~E.,
  Wolfe, B.~E., et~al. (2014).
\newblock Diet rapidly and reproducibly alters the human gut microbiome.
\newblock {\em Nature}, 505(7484):559--563.

\bibitem[Deutscher, 2008]{Deutscher2008}
Deutscher, J. (2008).
\newblock The mechanisms of carbon catabolite repression in bacteria.
\newblock {\em Current opinion in microbiology}, 11(2):87--93.

\bibitem[Dzierzawa e Om{\'e}ro, 2000]{Dzierzawa2000}
Dzierzawa, M. and Om{\'e}ro, M.-J. (2000).
\newblock Statistics of stable marriages.
\newblock {\em Physica A: Statistical Mechanics and its Applications},
  287(1):321--333.

\bibitem[Faith et~al., 2013]{Faith2013}
Faith, J.~J., Guruge, J.~L., Charbonneau, M., Subramanian, S., Seedorf, H.,
  Goodman, A.~L., et~al. (2013).
\newblock The long-term stability of the human gut microbiota.
\newblock {\em Science}, 341(6141):1237439.

\bibitem[Flint et~al., 2008]{Flint2008}
Flint, H.~J., Bayer, E.~A., Rincon, M.~T., Lamed, R., and White, B.~A. (2008).
\newblock Polysaccharide utilization by gut bacteria: potential for new
  insights from genomic analysis.
\newblock {\em Nature Reviews Microbiology}, 6(2):121--131.

\bibitem[Flynn et~al., 2017]{Antonopoulos2017}
Flynn, T.~M., Koval, J.~C., Greenwald, S.~M., Owens, S.~M., Kemner, K.~M., and
  Antonopoulos, D.~A. (2017).
\newblock Parallelized, aerobic, single carbon-source enrichments from
  different natural environments contain divergent microbial communities.
\newblock {\em Frontiers in Microbiology}, 8:2321.

\bibitem[Franzosa et~al., 2015]{Xochitl2015}
Franzosa, E.~A., Hsu, T., Sirota-Madi, A., Shafquat, A., Abu-Ali, G., Morgan,
  X.~C., et~al. (2015).
\newblock Sequencing and beyond: integrating molecular'omics' for microbial
  community profiling.
\newblock {\em Nature Reviews Microbiology}, 13(6):360--372.

\bibitem[Friedman et~al., 2017]{Friedman2017}
Friedman, J., Higgins, L.~M., and Gore, J. (2017).
\newblock Community structure follows simple assembly rules in microbial
  microcosms.
\newblock {\em Nature Ecology \& Evolution}, 1(5):s41559--017.

\bibitem[Fukami e Nakajima, 2011]{Fukami2011}
Fukami, T. and Nakajima, M. (2011).
\newblock Community assembly: alternative stable states or alternative
  transient states?
\newblock {\em Ecology letters}, 14(10):973--984.

\bibitem[Gale e Shapley, 1962]{Gale1962}
Gale, D. and Shapley, L.~S. (1962).
\newblock College admissions and the stability of marriage.
\newblock {\em The American Mathematical Monthly}, 69(1):9--15.

\bibitem[Goldford et~al., 2017]{Goldford2017}
Goldford, J.~E., Lu, N., Bajic, D., Estrela, S., Tikhonov, M.,
  Sanchez-Gorostiaga, A., et~al. (2017).
\newblock Emergent simplicity in microbial community assembly.
\newblock {\em bioRxiv}, pg. 205831.

\bibitem[Golubski et~al., 2016]{Golubski2016}
Golubski, A.~J., Westlund, E.~E., Vandermeer, J., and Pascual, M. (2016).
\newblock Ecological networks over the edge: hypergraph trait-mediated indirect
  interaction (tmii) structure.
\newblock {\em Trends in ecology \& evolution}, 31(5):344--354.

\bibitem[G{\"o}rke e St{\"u}lke, 2008]{Gorke2008}
G{\"o}rke, B. and St{\"u}lke, J. (2008).
\newblock Carbon catabolite repression in bacteria: many ways to make the most
  out of nutrients.
\newblock {\em Nature Reviews Microbiology}, 6(8):613--624.

\bibitem[Goyal e Maslov, 2018]{Goyal2017}
Goyal, A. and Maslov, S. (2018).
\newblock Diversity, stability, and reproducibility in stochastically assembled
  microbial ecosystems.
\newblock {\em Physical Review Letters}, 120(15):158102.

\bibitem[Graham et~al., 2007]{Graham2007}
Graham, D.~W., Knapp, C.~W., Van~Vleck, E.~S., Bloor, K., Lane, T.~B., and
  Graham, C.~E. (2007).
\newblock Experimental demonstration of chaotic instability in biological
  nitrification.
\newblock {\em The ISME journal}, 1(5):385--393.

\bibitem[Grilli et~al., 2017]{Grilli2017}
Grilli, J., Barab{\'a}s, G., Michalska-Smith, M.~J., and Allesina, S. (2017).
\newblock Higher-order interactions stabilize dynamics in competitive network
  models.
\newblock {\em Nature}, 548(7666):210--213.

\bibitem[Gusfield e Irving, 1989]{Gusfield1989}
Gusfield, D. and Irving, R.~W. (1989).
\newblock {\em The stable marriage problem: structure and algorithms}.
\newblock MIT press, Cambridge, MA.

\bibitem[Hardin et~al., 1960]{Hardin1960}
Hardin, G. et~al. (1960).
\newblock The competitive exclusion principle.
\newblock {\em science}, 131(3409):1292--1297.

\bibitem[Irving et~al., 2000]{Irving2000}
Irving, R.~W., Manlove, D.~F., and Scott, S. (2000).
\newblock The hospitals/residents problem with ties.
\newblock Em {\em SWAT}, pgs. 259--271. Springer.

\bibitem[Iwama e Miyazaki, 2015]{Miyazaki2015}
Iwama, K. and Miyazaki, S. (2015).
\newblock Stable marriage with ties and incomplete lists.
\newblock {\em Encyclopedia of Algorithms}, pgs. 883--885.

\bibitem[Klitgord e Segr{\`e}, 2010]{Klitgord2010}
Klitgord, N. and Segr{\`e}, D. (2010).
\newblock Environments that induce synthetic microbial ecosystems.
\newblock {\em PLoS computational biology}, 6(11):e1001002.

\bibitem[Knuth, 1997]{Knuth1997}
Knuth, D.~E. (1997).
\newblock {\em Stable marriage and its relation to other combinatorial
  problems: An introduction to the mathematical analysis of algorithms},
  volume~10.
\newblock American Mathematical Soc. Providence, RI.

\bibitem[Konopka, 2009]{Konopka2009}
Konopka, A. (2009).
\newblock What is microbial community ecology?
\newblock {\em The ISME journal}, 3(11):1223--1230.

\bibitem[Konopka et~al., 2015]{Konopka2015}
Konopka, A., Lindemann, S., and Fredrickson, J. (2015).
\newblock Dynamics in microbial communities: unraveling mechanisms to identify
  principles.
\newblock {\em The ISME journal}, 9(7):1488--1495.

\bibitem[Koropatkin et~al., 2012]{Martens2012}
Koropatkin, N.~M., Cameron, E.~A., and Martens, E.~C. (2012).
\newblock How glycan metabolism shapes the human gut microbiota.
\newblock {\em Nature Reviews Microbiology}, 10(5):323--335.

\bibitem[Kremling et~al., 2015]{Kremling2015}
Kremling, A., Geiselmann, J., Ropers, D., and De~Jong, H. (2015).
\newblock Understanding carbon catabolite repression in escherichia coli using
  quantitative models.
\newblock {\em Trends in microbiology}, 23(2):99--109.

\bibitem[Levine et~al., 2017]{Levine2017}
Levine, J.~M., Bascompte, J., Adler, P.~B., and Allesina, S. (2017).
\newblock Beyond pairwise mechanisms of species coexistence in complex
  communities.
\newblock {\em Nature}, 546(7656):56--64.

\bibitem[Levy e Borenstein, 2013]{Levy2013}
Levy, R. and Borenstein, E. (2013).
\newblock Metabolic modeling of species interaction in the human microbiome
  elucidates community-level assembly rules.
\newblock {\em Proceedings of the National Academy of Sciences},
  110(31):12804--12809.

\bibitem[Lozupone et~al., 2012]{Lozupone2012}
Lozupone, C.~A., Stombaugh, J.~I., Gordon, J.~I., Jansson, J.~K., and Knight,
  R. (2012).
\newblock Diversity, stability and resilience of the human gut microbiota.
\newblock {\em Nature}, 489(7415):220--230.

\bibitem[MacArthur e MacArthur, 1961]{MacArthur1961}
MacArthur, R.~H. and MacArthur, J.~W. (1961).
\newblock On bird species diversity.
\newblock {\em Ecology}, 42(3):594--598.

\bibitem[Manlove et~al., 2002]{Manlove2002}
Manlove, D.~F., Irving, R.~W., Iwama, K., Miyazaki, S., and Morita, Y. (2002).
\newblock Hard variants of stable marriage.
\newblock {\em Theoretical Computer Science}, 276(1-2):261--279.

\bibitem[McVitie e Wilson, 1971]{McVitie1971}
McVitie, D.~G. and Wilson, L.~B. (1971).
\newblock The stable marriage problem.
\newblock {\em Communications of the ACM}, 14(7):486--490.

\bibitem[Momeni et~al., 2017]{Momeni2017}
Momeni, B., Xie, L., and Shou, W. (2017).
\newblock Lotka-volterra pairwise modeling fails to capture diverse pairwise
  microbial interactions.
\newblock {\em Elife}, 6.

\bibitem[Monod, 1949]{Monod1949}
Monod, J. (1949).
\newblock The growth of bacterial cultures.
\newblock {\em Annual Reviews in Microbiology}, 3(1):371--394.

\bibitem[Posfai et~al., 2017]{Wingreen2017}
Posfai, A., Taillefumier, T., and Wingreen, N.~S. (2017).
\newblock Metabolic trade-offs promote diversity in a model ecosystem.
\newblock {\em Physical Review Letters}, 118(2):028103.

\bibitem[Qin et~al., 2010]{Qin2010}
Qin, J., Li, R., Raes, J., Arumugam, M., Burgdorf, K.~S., Manichanh, C., et~al.
  (2010).
\newblock A human gut microbial gene catalogue established by metagenomic
  sequencing.
\newblock {\em nature}, 464(7285):59--65.

\bibitem[Raghavan e Groisman, 2015]{Raghavan2015}
Raghavan, V. and Groisman, E.~A. (2015).
\newblock Species-specific dynamic responses of gut bacteria to a mammalian
  glycan.
\newblock {\em Journal of bacteriology}, 197(9):1538--1548.

\bibitem[Rakoff-Nahoum et~al., 2014]{Comstock2015}
Rakoff-Nahoum, S., Coyne, M.~J., and Comstock, L.~E. (2014).
\newblock An ecological network of polysaccharide utilization among human
  intestinal symbionts.
\newblock {\em Current Biology}, 24(1):40--49.

\bibitem[Rivi{\`e}re et~al., 2018]{Riviere2018}
Rivi{\`e}re, A., Selak, M., Geirnaert, A., Van~den Abbeele, P., and De~Vuyst,
  L. (2018).
\newblock Complementary mechanisms for degradation of inulin-type fructans and
  arabinoxylan oligosaccharides among bifidobacterial strains suggest bacterial
  cooperation.
\newblock {\em Applied and environmental microbiology}, 84(9):e02893--17.

\bibitem[Rogowski et~al., 2015]{Martens2015}
Rogowski, A., Briggs, J.~A., Mortimer, J.~C., Tryfona, T., Terrapon, N., Lowe,
  E.~C., et~al. (2015).
\newblock Glycan complexity dictates microbial resource allocation in the large
  intestine.
\newblock {\em Nature communications}, 6.

\bibitem[Roth e Sotomayor, 1992]{Roth1992}
Roth, A.~E. and Sotomayor, M. (1992).
\newblock Two-sided matching.
\newblock {\em Handbook of game theory with economic applications}, 1:485--541.

\bibitem[Skupin e Metzger, 2015]{Skupin2015}
Skupin, P. and Metzger, M. (2015).
\newblock Oscillatory behavior control in continuous fermentation processes.
\newblock {\em IFAC-PapersOnLine}, 48(8):1114--1119.

\bibitem[Sung et~al., 2017]{Chia2017}
Sung, J., Kim, S., Cabatbat, J., Jang, S., Jin, Y., Jung, G., et~al. (2017).
\newblock Global metabolic interaction network of the human gut microbiota for
  context-specific community-scale analysis.
\newblock {\em Nature communications}, 8:15393.

\bibitem[Tikhonov e Monasson, 2017]{Tikhonov2017}
Tikhonov, M. and Monasson, R. (2017).
\newblock Collective phase in resource competition in a highly diverse
  ecosystem.
\newblock {\em Physical Review Letters}, 118(4):048103.

\bibitem[Tilman, 1982]{Tilman1982}
Tilman, D. (1982).
\newblock {\em Resource competition and community structure}.
\newblock Princeton university press.

\bibitem[Tuncil et~al., 2017]{Martens2017}
Tuncil, Y.~E., Xiao, Y., Porter, N.~T., Reuhs, B.~L., Martens, E.~C., and
  Hamaker, B.~R. (2017).
\newblock Reciprocal prioritization to dietary glycans by gut bacteria in a
  competitive environment promotes stable coexistence.
\newblock {\em mBio}, 8(5):e01068--17.

\bibitem[Venturelli et~al., 2017]{Venturelli2017}
Venturelli, O.~S., Carr, A.~C., Fisher, G., Hsu, R.~H., Lau, R., Bowen, B.~P.,
  et~al. (2017).
\newblock Deciphering microbial interactions in synthetic human gut microbiome
  communities.
\newblock {\em bioRxiv}, pg. 228395.

\bibitem[Wang e Tang, 2017]{Wang2017}
Wang, X. and Tang, C. (2017).
\newblock Optimal growth of microbes on mixed carbon sources.
\newblock {\em arXiv preprint arXiv:1703.08791}.

\bibitem[Weiher et~al., 1998]{Weiher1998}
Weiher, E., Clarke, G.~P., and Keddy, P.~A. (1998).
\newblock Community assembly rules, morphological dispersion, and the
  coexistence of plant species.
\newblock {\em Oikos}, pgs. 309--322.

\bibitem[Zamamiri et~al., 2001]{Zamamiri2001}
Zamamiri, A.-Q.~M., Birol, G., and Hjorts{\o}, M.~A. (2001).
\newblock Multiple stable states and hysteresis in continuous, oscillating
  cultures of budding yeast.
\newblock {\em Biotechnology and bioengineering}, 75(3):305--312.

\bibitem[Zhang, 2001]{Zhang2001}
Zhang, Y.-C. (2001).
\newblock Happier world with more information.
\newblock {\em Physica A: Statistical Mechanics and its Applications},
  299(1):104--120.

\bibitem[Zhou et~al., 2013]{Zhou2013}
Zhou, J., Liu, W., Deng, Y., Jiang, Y.-H., Xue, K., He, Z., et~al. (2013).
\newblock Stochastic assembly leads to alternative communities with distinct
  functions in a bioreactor microbial community.
\newblock {\em Mbio}, 4(2):e00584--12.

\end{thebibliography}

\clearpage
\section*{Supplementary Information}
\renewcommand{\thefigure}{S\arabic{figure}}
\setcounter{figure}{0}
\vfill
\begin{figure}[h]
\begin{center}
\includegraphics[width=0.95\linewidth]{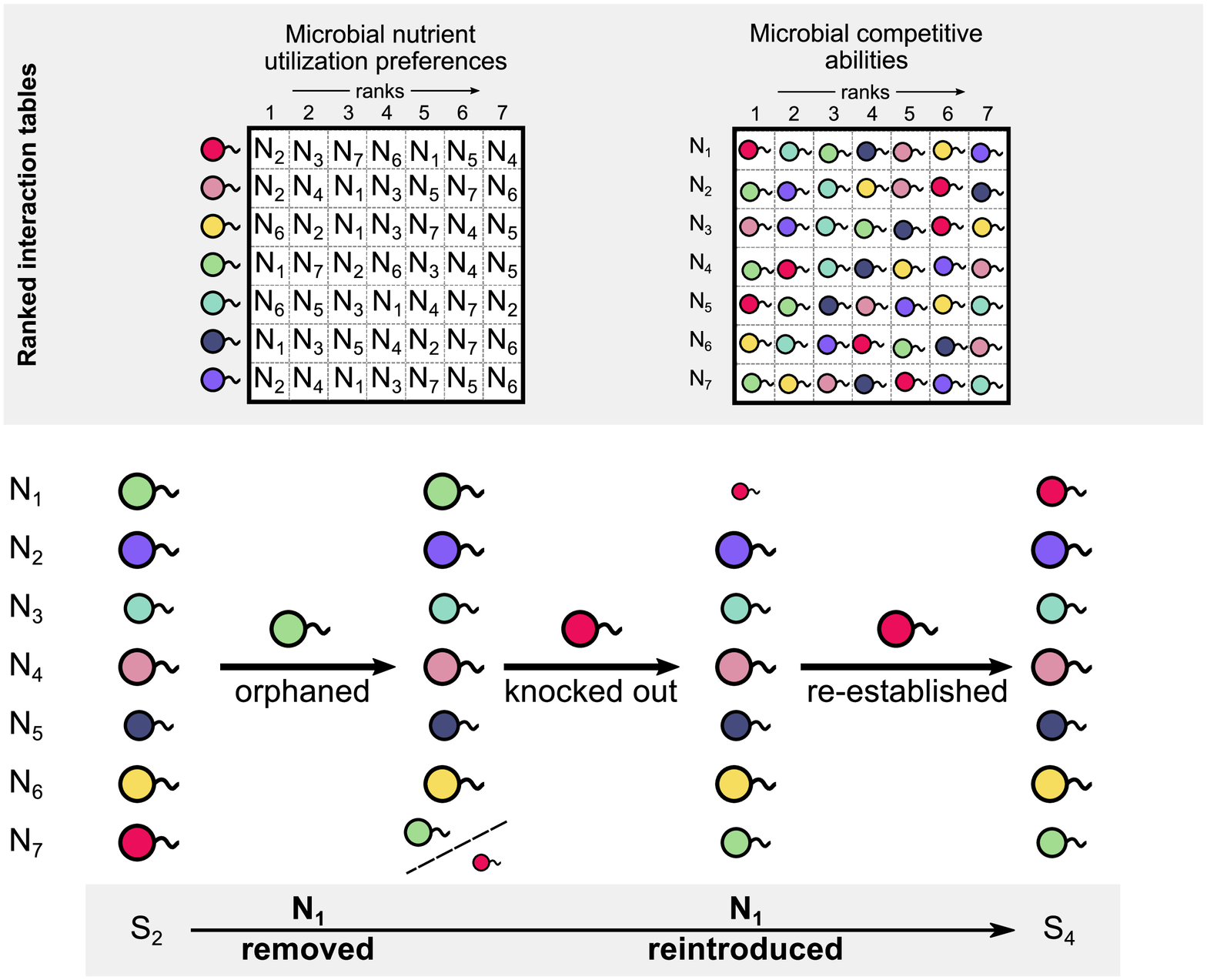}
\caption{\textbf{Specific steps during community transitions from one stable state to another.} \\
A detailed step-by-step breakdown of how the microbial community in our example in figure \ref{keyfig2} of the main text can transition from one stable state (here $S_2$) to another (here $S_4$) via a very specific perturbation: the removal of nutrient $N_1$ and its reintroduction at the specific time-point shown thereafter. First, the green microbe is left without its preferred growth nutrient ($N_1$). It then attempts to compete for its next preferred nutrient, $N_7$, competitively displaces the red microbe, which can then re-establish on $N_1$ reintroduced at that specific time. The resultant community now exhibits the alternate stable state, $S_4$.}
\label{suppfig2}
\end{center}
\end{figure}
\vfill

\clearpage
\vfill
\begin{figure}
\begin{center}
\includegraphics[width=0.95\linewidth]{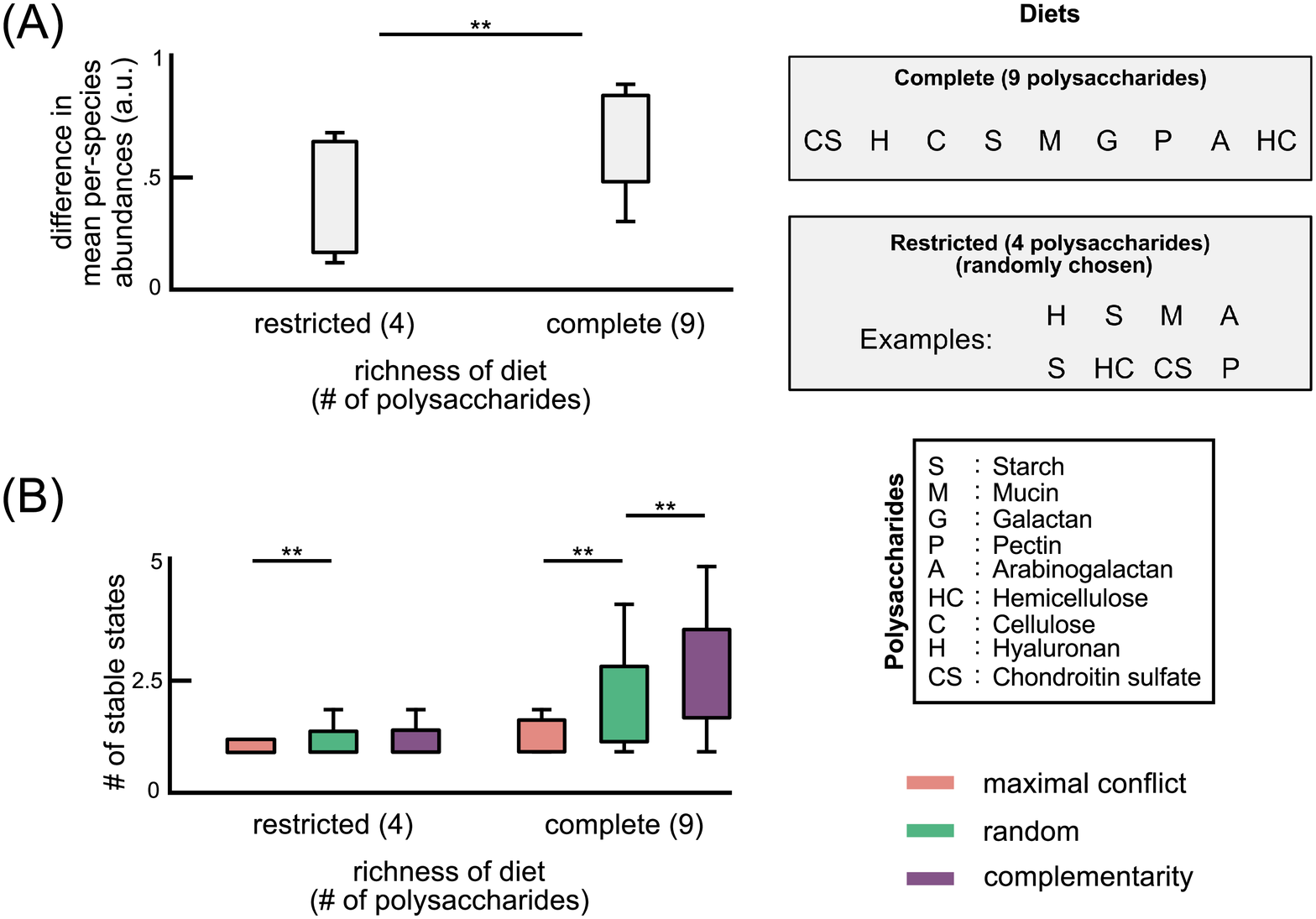}
\caption{\textbf{Contrast between restricted and complete diets in \emph{Bacteroides} species.} \\
\textbf{(A)} In figure \ref{keyfig3} in the main text, we show that using different nutrient preferences (complementary, random and maximally conflicting) for a realistic community of \emph{Bacteroides} species can result in different species abundance profiles. Specifically, complementary lists lead to higher abundances for all \emph{Bacteroides} species, whereas conflicting lists result in low abundances. However, we showed this assuming a complete `diet' with all 9 consumable polysaccharides available. Here, we show that the difference between communities with complementary and conflicting preferences (in our model) shrinks when the diet is `restricted', i.e. when only about half the polysaccharides are available, and randomly selected. This is consistent with an increased expectation for complementary nutrient preferences between co-occurring microbes in environments with richer diets. \textbf{(B)} The number of stable states, as described in the main text (see Materials and methods: Enumerating all stable states) for all three cases of microbial nutrient preferences for restricted and complete diets. Complete diets typically have a higher number of stable states (typically $\sim 2$) for complementary preferences than either random or conflicting preferences. In some cases, the number of stable states is higher (i.e. $4-5$), and these cases are more likely if the preferences are complementary. (In all cases, we use the Kolmogorov-Smirnov test to compare distributions, with a $P$-value threshold of $0.01$.) 
}
\label{suppfig3}
\end{center}
\end{figure}
\vfill

\clearpage
\vfill
\begin{figure}
\begin{center}
\includegraphics[width=0.95\linewidth]{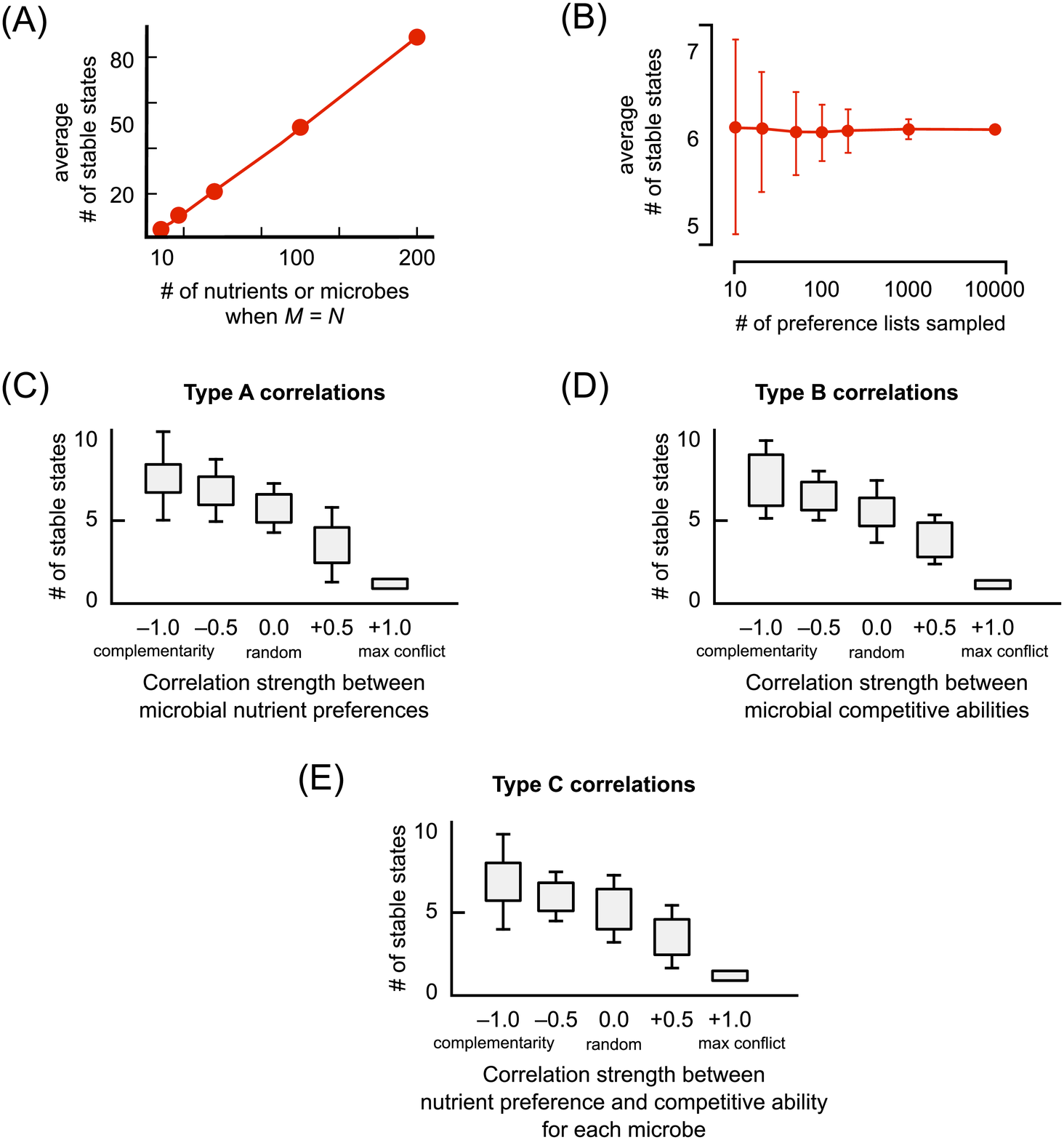}
\caption{\textbf{Characteristics of the number of stable states according to our model} \\
\textbf{(A)} The average number of stable states for $1,000$ randomized preference lists (when $M=N$ on the x-axis, $M$ and $N$ being the number of microbes and nutrients respectively). \textbf{(B)} For $M=N=10$, this plot shows the average number of stable states using randomized preference lists of different sizes (on the x-axis). The error-bars represent the standard deviation among different samples of the same size. \textbf{(C, D, E)} Box plots of the number of stable states for different type A (microbes' nutrient preferences), type B (microbes' competitive abilities for a nutrient) and type C (preference for a nutrient and competitive ability on it for a particular microbe) correlations (here $M=N=10$). Correlations strengths are shown on the x-axis. They can either be complementary (anti-correlated preferences, or strength $-1$), random (uncorrelated preferences, or strength $0$), or maximally conflicting (strongly positively correlated preferences, or strength $+1$). For all types of correlations, strong positive correlations result in a community with only one unique stable state.
}
\label{suppfig3}
\end{center}
\end{figure}
\vfill
\clearpage
\section*{Supplementary Methods}
\subsection*{Studying correlations between preference lists}
We assumed a microbial community with 10 microbes and 10 nutrients (i.e., $M=N=10$). For each community, we generated a unique set of preferences that were correlated according to a chosen strength $x$. For each microbial species $i$ on each nutrient $\alpha$, we first generated two numbers: one indicating the microbes preference for that nutrient, and the other for its competitive ability to uptake that nutrient given competition with others. In other words, we generated ``preference values'' $P_{i\alpha}$ and ``competitive ability values'' $C_{i\alpha}$ for each microbe $i$ on each nutrient $\alpha$ such that these values reflected our desired correlations of strength $x$. 

After generating these values, we rank ordered the nutrient preferences and competitive abilities to generate the two ranked tables that fully specify our stable marriage model, and used our previously described algorithm to enumerate the number of stable states (see Methods in the main text). Below we describe our method to generate three types of correlations (see Discussion). 

First, type A correlations indicate correlations between the nutrient preferences of all microbes, i.e. they reflect a case when certain nutrients are ``universally prized'' or preferred. Here, we randomized all competitive ability values $C_{i\alpha}$'s from a uniform distribution between 0 and 1. For nutrient preferences, we assigned $P_{i\alpha}$'s in the following way.

\begin{align}
P_{i\alpha} &= x \lambda \alpha + \sqrt{1 - x^2} \lambda'_{i\alpha} \hskip 20pt \forall \hskip 20pt 0 \leq x \leq 1.
\end{align}

All $\lambda$'s represent random numbers distributed uniformly between 0 and 1. Note that for each $i$ and $\alpha$, we generate a unique $\lambda'$, but a shared $\lambda$. Further, here $0 \leq x \leq 1$ represents positively correlated (conflicting) preferences. For $1 \leq x < 0$ which represents negatively correlated (complementary) preferences, we used our original procedure to generate complementary nutrient preferences (see Methods in the main text), with the following change. At $x = -1$ (maximum complementarity), we always assigned a unique nutrient as a microbe's top choice, but at $x = -0.5$, we assigned a unique nutrient with probability $\lvert x \rvert = 0.5$.

Second, type B correlations indicate correlations between the competitive abilities of all microbes on the same nutrient. Here, we randomized microbes nutrient preferences $P_{i\alpha}$'s and used a similar algorithm for the type A correlations above, except with competitive abilities $C_{i\alpha}$'s, as follows.

\begin{align}
C_{i\alpha} &= x \lambda \alpha + \sqrt{1 - x^2} \lambda'_{i\alpha} \hskip 20pt \forall \hskip 20pt 0 \leq x \leq 1.
\end{align}

Finally, type C correlations indicate correlations between a microbe's preference for a nutrient and its competitive ability on that nutrient. For this, for each microbe $i$ and nutrient $\alpha$, we generate both the preference and competitive ability values using the following equations.

\begin{align}
P_{i\alpha} &= \lambda_{i\alpha} + \sqrt{1 - x^2} \lambda'_{i\alpha} \\
C_{i\alpha} &= x \lambda_{i\alpha} + \sqrt{1 - x^2} \lambda''_{i\alpha}.
\end{align}

\end{document}